\documentclass[english]{article}
\usepackage[T1]{fontenc}
\usepackage[latin9]{inputenc}
\usepackage{geometry}
\usepackage{float}
\usepackage{amsthm}
\usepackage{amsmath}
\usepackage{graphicx}
\usepackage{setspace}
\usepackage[numbers]{natbib}
\makeatletter

\providecommand{\tabularnewline}{\\}
\newcommand{\lyxdot}{.}

\newcommand{\lyxaddress}[1]{
\par {\raggedright #1
\vspace{1.4em}
\noindent\par}
}

\makeatother

\usepackage{babel}
\begin{document}

\title{Empirical Bayes estimation of posterior probabilities of enrichment}

\author{Zhenyu Yang$^{1}$, Zuojing Li$^{2}$ and David R. Bickel$^{1*}$}

\maketitle

\lyxaddress{$^{1}$Ottawa Institute of Systems Biology, Department of Biochemistry,
Microbiology, and Immunology, University of Ottawa, 451 Smyth Road,
Ottawa, Ontario, Canada, K1H 8M5}

\lyxaddress{$^{2}$School of Foundation, Shenyang Pharmaceutical University,
No. 103 Wenhua Road, Shenyang, Liaoning, 110016, China}

\lyxaddress{E-mail: zyang009@uottawa.ca; zuojing1006@hotmail.com; dbickel@uottawa.ca}

\lyxaddress{$^{*}$Corresponding author}
\begin{abstract}
\textbf{Background:} To interpret differentially expressed genes or
other discovered features, researchers conduct hypothesis tests to
determine which biological categories such as those of the Gene Ontology
(GO) are enriched in the sense of having differential representation
among the discovered features. Multiple comparison procedures (MCPs)
are commonly used to prevent excessive false positive rates. Traditional
MCPs, e.g., the Bonferroni correction, go to the opposite extreme
of strictly controlling a family-wise error rate, resulting in excessive
false negative rates. Researchers generally prefer the more balanced
approach of instead controlling the false discovery rate (FDR). Methods
of FDR control assign q-values to biological categories, but q-values
are too low to reliably estimate a probability that the biological
category has equivalent representation among the preselected features.
Thus, we study application of better estimators of that probability,
which is technically known as the local false discovery rate (LFDR).

\textbf{Results:} We identified three promising estimators of the
LFDR for detecting differential representation: a semiparametric estimator
(SPE), a\emph{ }normalized maximum likelihood estimator (NMLE), and
a maximum likelihood estimator (MLE). We found that the MLE performs
at least as well as the SPE for on the order of 100 of GO categories
even when the ideal number of components in its underlying mixture
model is unknown. However, the MLE is unreliable when the number of
GO categories is small compared to the number of PMM components. Thus,
if the number of categories is on the order of 10, the SPE is a more
reliable LFDR estimator. The NMLE depends not only on the data but
also on a specified value of the prior probability of differential
representation. It is therefore an appropriate LFDR estimator only
when the number of GO categories is too small for application of the
other methods.

\textbf{Conclusions}: For enrichment detection, we recommend estimating
the LFDR by the MLE given at least a medium number ($\sim$100) of
GO categories, by the SPE given a small number of GO categories ($\sim$10),
and by the NMLE given a very small number ($\sim$1) of GO categories.
\end{abstract}
\newpage{}

\section*{Introduction\label{sec:Introduction}}

The development of microarray techniques and high-throughput genomic,
proteomic, and bioinformatics scanning approaches (such as microarray
gene expression profiling, mass spectrometry, and ChIP-on-chip) has
enabled researchers simultaneously to study tens of thousands of biological
features (e.g., genes, proteins, single-nucleotide polymorphisms {[}SNPs{]},
etc.), and to identify a set of features for further investigation.
However, there remains the challenge of interpreting these features
biologically. For a given set of features, the determination of whether
some biological information terms are differentially represented (i.e.,
overrepresented or underrepresented), compared to the reference feature
set, is termed the \emph{feature enrichment} problem. The biological
information term may be, for instance, a Gene Ontology (GO) category
\citep{altshuler_genetic_2008} or a pathway in the Kyoto Encyclopedia
of Genes and Genomes (KEGG) \citep{kanehisa_kegg_2000}. 

This problem has been addressed using a number of high-throughput
enrichment tools, including DAVID \citep{dennis_david_2003}, MAPPFinder
\citep{doniger_mappfinder_2003}, Onto-Express \citep{khatri_profiling_2002},
and GoMiner \citep{zeeberg_gominer_2003}. \citet{huang_bioinformatics_2009}
reviewed $68$ distinct feature enrichment analysis tools. These authors
further classified feature enrichment analysis tools into $3$ categories:
singular enrichment analysis (SEA), gene set enrichment analysis (GSEA)
and modular enrichment analysis (MEA). Here, we investigate the SEA
problem using gene expression as a concrete example. More precisely,
we consider whether some specific biological categories are differentially
represented among the preselected genes with respect to the reference
genes. We call this problem the \emph{gene enrichment problem}. 

Existing enrichment tools mainly address the gene enrichment problem
using a p-value obtained from an exact or approximate statistical
test (e.g., Fisher's exact test, or the hypergeometric test, binomial
test, $\chi^{2}$ test, etc.). For each GO term or other biological
category, the null hypothesis tested and its alternative hypothesis
are these:

\begin{equation}
\begin{array}{c}
H_{0}:\text{ the GO category is equivalently represented among the preselected genes}\text{ }\\
H_{1}:\text{ the GO category is differentially represented among the preselected genes}
\end{array}\label{eqn:Hypothesis-comparison}
\end{equation}
The general process begins as follows: 
\begin{itemize}
\item For each GO category, construct Table \ref{tab:twobytwogenes} based
on the preselected genes (e.g., differentially expressed (DE) genes)
and reference genes (e.g., all genes measured in a microarray experiment). 
\item Compute the p-value for each GO category using a statistical test
that can detect enrichment in the sense of differential representation
among the preselected genes.
\end{itemize}
\begin{table}[tbph]
\caption{The number of differentially expressed (DE) and equivalently expressed
(EE) genes in a GO category. Here, $x_{i}$ ($i=1,2$) is the number
of DE ($i=1$) or EE genes ($i=2$) in the GO category; $n$ is the
total number of DE genes; $N$ is the total number of reference genes.\label{tab:twobytwogenes}}

\centering{}%
\begin{tabular}{lccc}
\hline 
 & DE genes & EE gene & Total\tabularnewline
\hline 
In GO category & $x_{1}$ & $x_{2}$ & $x_{1}+x_{2}$\tabularnewline
Not in GO category & $n-x_{1}$ & $N-n-x_{2}$  & $N-x_{1}-x_{2}$\tabularnewline
Total & $n$ & $N-n$ & $N$\tabularnewline
\hline 
\end{tabular}
\end{table}

Multiple comparison procedures (MCPs) are then applied to the resulting
p-values to prevent excessive false-positive rates. The false discovery
rate (FDR) \citep{RefWorks:288} is frequently used to control the
expected proportion of incorrectly rejected null hypotheses in gene
enrichment studies \citep{Min_Variability_2010,Reyal_comprehensive_2008,wang_transcriptional_2011}
because it has lower false-negative rates than the Bonferroni correction
and other methods of controlling the family-wise error rate. Methods
of FDR control assign q-values \citep{storey_qvalue_2003} to biological
categories, but q-values are too low to reliably estimate the probability
that the biological category has equivalent representation among the
preselected features. Thus, we study application of better estimators
of that probability, which is technically known as the local false
discovery rate (LFDR). \citet{Hong_lfdr_2009} used an LFDR estimator
to solve a GSEA problem and pointed out that this was less biased
than the q-value for estimating the LFDR, the posterior probability
that the null hypothesis is true. 

\citet{RefWorks:55,efron_large-scale_2010} devised reliable LFDR
estimators for a range of applications in microarray gene expression
analysis and other problems of large-scale inference. However, whereas
microarray gene expression analysis takes into account tens of thousands
of genes, the gene enrichment problem typically concerns a much smaller
number of GO categories. While those methods are appropriate for microarray-scale
inference, they are less reliable for enrichment-scale inference \citet{mediumScale,smallScale}.
Thus, we will specifically adapt three types of LFDR estimators that
are appropriate for smaller-scale inference to address the SEA problem.
Here we will focus on genes and GO categories. Nevertheless, the estimators
used can be broadly applied to other features (e.g., proteins, SNPs)
and biological terms (e.g., those featuring metabolic pathways). 

The sections of this paper are arranged as follows. We will first
introduce some preliminary concepts in the gene enrichment problem.
Next, $3$ LFDR estimators will be described. After that, we will
compare the LFDR estimators using breast cancer data and simulation
data. Finally, we will draw conclusions and make recommendations on
the basis of our results.

\section*{Preliminary concepts\label{sec:preconcepts}}

The gene enrichment problem described in the Introduction is stated
here more formally for application of LFDR methods of the next section.

\subsection*{Likelihood functions\label{sub:LikelihoodFun} }

Consider Table \ref{tab:twobytwogenes}. Let $X_{1}$ and $X_{2}$
respectively denote the random numbers of DE and EE genes in a GO
category. The resulting categories follow the binomial distribution,
i.e.,$X_{1}\sim\text{Binomial}(n,\Pi_{1})$ and $X_{2}\sim\text{Binomial}(N-n,\Pi_{2})$,
where $\Pi_{1}$ is the proportion of DE genes in the GO category
and $\Pi_{2}$ is the proportion of EE genes in the GO category. Under
the assumption that $X_{1}$ and $X_{2}$ are independent, the\emph{
unconditional likelihood} is 
\begin{eqnarray}
 &  & L(\Pi_{1},\Pi_{2};x_{1},x_{2},n,N)\label{eqa:FullLikelihood}\\
 &  & =\text{Pr}(X_{1}=x_{1},X_{2}=x_{2};\Pi_{1},\Pi_{2},n,N)\nonumber \\
 &  & =\binom{n}{x_{1}}\binom{N-n}{x_{2}}\Pi_{1}^{x_{1}}(1-\Pi_{1})^{n-x_{1}}\Pi_{2}^{x_{2}}(1-\Pi_{2})^{N-n-x_{2}}\nonumber 
\end{eqnarray}
where $0\le x_{1}\le n$, $0\le x_{2}\le N-n$ and $0\le\Pi_{i}\le1$,
$i=1,2$.

If we define 
\begin{equation}
\lambda=\ln[\Pi_{2}/(1-\Pi_{2})]\label{eqn:lambda}
\end{equation}
 and 
\begin{equation}
\theta=\ln[\Pi_{1}/(1-\Pi_{1})]-\lambda\label{eqn:theta}
\end{equation}
then $\theta$ is the parameter of interest, representing the \emph{log
odds ratio} of the GO category, and $\lambda$ is a nuisance parameter.
Under the new parametrization, the unconditional likelihood function
\eqref{eqa:FullLikelihood} is 

\begin{equation}
L(\theta,\lambda;x_{1},x_{2},n,N)=\frac{\binom{n}{x_{1}}\binom{N-n}{x_{2}}\times e^{x_{1}(\theta+\lambda)}e^{x_{2}\lambda}}{(1+e^{\theta+\lambda})^{n_{1}}(1+e^{\lambda})^{n_{2}}}\label{eqn:FullLikelihoodNew}
\end{equation}
where $0\le x_{1}\le n$ and $0\le x_{2}\le N-n$.

In equation \eqref{eqn:FullLikelihoodNew}, we take the interest parameter
$\theta$ and also the nuisance parameter $\lambda$ into consideration.
Consider statistics $T$ and $S$, functions of $X_{1}$ and $X_{2}$,
such that $T(X_{1},X_{2})=X_{1}$ and $S(X_{1},X_{2})=X_{1}+X_{2}$.
Thus, $T$ represents the number of DE genes in a GO category, and
$S$ represents the number of total genes in a GO category. Let $t$
and $s$ be the observed values of statistic $T$ and $S$. The probability
mass function of $T(x_{1},x_{2})=t$ evaluated at $S(x_{1},x_{2})=x_{1}+x_{2}=s$,
say $\text{Pr}(T=t|S=s;\theta,\lambda,N,n)$, does not depend on the
nuisance parameter $\lambda$ \citep{mediumScale}. See also Example
$8.47$ of \citet{Severini2000}. Thus, we derive the conditional
probability mass function 
\begin{equation}
f_{\theta}(t|s)=\text{Pr}(T=t|S=s;\theta,n,N)=\frac{\binom{n}{t}\binom{N-n}{v-t}e^{t\theta}}{\sum_{j=\max(0,s+n-N)}^{\min(s,n)}\binom{n}{j}\binom{N-n}{s-j}e^{j\theta}}\label{eqn:Condmassfunc}
\end{equation}
understood as a function of $t$.

By eliminating the nuisance parameter $\lambda$, we can reduce the
original data $x_{1}$ and $x_{2}$ by considering the statistic $T=t$.
However, the use of the conditional probability mass function requires
some justification because of concerns about losing information during
the conditioning process. Unfortunately, in the presence of the nuisance
parameter, the statistic $S(X_{1},X_{2})=X_{1}+X_{2}$ is not an ancillary
statistic for the parameter of interest. In other words, the probability
mass function of the conditional variable $S(X_{1},X_{2})$ may contain
some information about the parameter $\theta$ \citep{Severini2000}.
However, following the explanation of \citet[$\mathcal{x} 2.5$]{RefWorks:436},
the expectation value of the statistic $S(X_{1},X_{2})$ equals the
nuisance parameter. Hence, from the observation of $S(X_{1},X_{2})$
alone, the distribution of the statistic $S(X_{1},X_{2})$ contains
little information about the parameter $\theta$ \citep{RefWorks:436}.
The statistic $S(X_{1},X_{2})$ satisfies the other $3$ conditions
of an ancillary statistic defined by \citet{RefWorks:436}: parameters
$\theta$ and $\lambda$ are variation independent; the statistic
($T(X_{1},X_{2}),S(X_{1},X_{2})$) is the minimal sufficient statistic;
and the distribution of the statistic $T(X_{1},X_{2})$, given $S(X_{1},X_{2})=s$,
is independent of parameter of interest, $\theta$, given the nuisance
parameter $\lambda$. Therefore, the probability mass function of
the statistic $S(X_{1},X_{2})$ contains little information about
the value of the parameter $\theta$.

\subsection*{Hypotheses and false discovery rates\label{sub:Hypothesis-comparisons}}

Considering GO category $i$, we denote the $T$, $S$, $t$, $s$
and $\theta$ used in equation \eqref{eqn:Condmassfunc} as $T_{i}$,
$S_{i}$, $t_{i}$, $s_{i}$ and $\theta_{i}$. From Table \ref{tab:twobytwogenes},
the hypothesis comparison \eqref{eqn:Hypothesis-comparison} of GO
category $i$ is equivalent to 

\begin{equation}
\theta_{i}=0\text{ versus }\theta_{i}\neq0\label{eqn:two-sided-hypothesis-GO}
\end{equation}
Let $\mathbf{S}=\left\langle S_{1},S_{2},\cdots,S_{m}\right\rangle $
and $\mathbf{s}=\left\langle s_{1},s_{2},\cdots,s_{m}\right\rangle $.
Let $\text{BF}_{i}$ denote the \emph{Bayes factor }of GO category
$i$:
\begin{equation}
\text{BF}_{i}=\frac{\text{Pr}(T_{i}=t_{i}|\mathbf{S}=\mathbf{s},\theta_{i}\ne0)}{\text{Pr}(T_{i}=t_{i}|\mathbf{S}=\mathbf{s},\theta_{i}=0)}\label{eqn:BF}
\end{equation}
It is called the Bayes factor because it yields the posterior odds
when multiplied by the prior odds. More precisely, the \emph{posterior
odds }of the alternative hypothesis corresponding to GO category $i$
is 
\begin{equation}
\omega_{i}=\frac{\text{Pr}(\theta_{i}\ne0|t_{i})}{\text{Pr}(\theta_{i}=0|t_{i})}=\text{BF}_{i}\times\frac{(1-\pi_{0})}{\pi_{0}}\label{eqn:OR}
\end{equation}
where $\pi_{0}$ is the \emph{prior conditional probability} that
a GO category is equivalently represented among the preselected genes
given $\mathbf{s}$, i.e., $\pi_{0}=\text{Pr}(\theta_{i}=0|\mathbf{S}=\mathbf{s})$.
Thus, $(1-\pi_{0})/\pi_{0}$ is the \emph{prior odds }of the alternative
hypothesis\emph{ }of differential representation\emph{. }According
to Bayes' theorem, the LFDR of GO category $i$ is

\begin{equation}
\text{LFDR}_{i}=\text{Pr}(\theta_{i}=0|t_{i})=\frac{1}{1+\omega_{i}},\label{eqn:LFDR-computation}
\end{equation}
where $\omega_{i}$ is defined in equation \eqref{eqn:OR}.

\section*{LFDR estimation methods\label{sec:Methods}}

\subsection*{Semiparametric LFDR estimator\label{sub:SPE}}

Let $\alpha$ denote any significance level chosen to be between 0
and 1. For all GO categories of interest, the FDR may be estimated
by 
\begin{eqnarray}
\widehat{\text{FDR}}(\alpha) & = & \min\left(\frac{m\alpha}{\sum_{j=1}^{m}\mathbf{1}_{\{p_{j}\le\alpha\}}},1\right)\label{eqn:FDR}
\end{eqnarray}
where $m$ is the number of GO categories, $p_{j}$ is the p-value
of GO category $j$, and $\mathbf{1}_{\{p_{j}\le\alpha\}}$ is the
indicator such that $\mathbf{1}_{\{p_{j}\le\alpha\}}=1$ if $p_{j}\le\alpha$
is true and $\mathbf{1}_{\{p_{j}\le\alpha\}}=0$ otherwise. Thus,
$\sum_{j=1}^{m}\mathbf{1}_{\{p_{j}\le\alpha\}}$ represents the number
of GO categories with discovered differential representation, and
$m\alpha$ estimates the number of such discoveries that are false. 

Let $r_{i}$ be the rank of the p-value of GO category $i$, e.g.,
$r_{i}=1$ if the p-value of GO category $i$ is the smallest among
all p-values of $m$ GO categories. Based on a modification of equation
\eqref{eqn:FDR}, the \emph{semiparametric estimator }(SPE) of LFDR
of the GO category $i$ is

\begin{equation}
\widehat{\text{LFDR}}_{i}=\begin{cases}
\min\left(\frac{mp_{2r_{i}}}{2r_{i}},1\right), & r_{i}\le\frac{m}{2}\\
1, & r_{i}>\frac{m}{2}
\end{cases}\label{eqn:r-value}
\end{equation}
It is conservative in the sense that it tends to overestimate the
LFDR \citep{BFDR}.

\subsection*{Type II maximum likelihood estimator\label{sec:MLE}}

\citet{BFDR} follows \citet{GOOD01121966} in calling the maximization
of likelihood over a hyperparameter \emph{Type II maximum likelihood}
to distinguish it from the usual \emph{Type I maximum likelihood},
which pertains only to models that lack random parameters. Type II
maximum likelihood has been applied to parametric mixture models for
the analysis of microarray data \citep{Pawitan20053865,ParametricMixtureLFDR},
proteomics data \citep{smallScale}, and genetic association data
\citep{GWAselect}. In this section, we adapt the approach to the
gene enrichment problem by using the conditional probability mass
function defined above.

Let $\mathcal{G}(\mathbf{s})=\{g_{\theta}(\bullet|\mathbf{s});\theta\ge0\}$
be a parametric family of probability mass functions with 
\begin{eqnarray}
g_{\theta}(\bullet|\mathbf{s}) & = & \frac{1}{2}\times\left[f_{\theta}(\bullet|\mathbf{s})+f_{-\theta}(\bullet|\mathbf{s})\right]\label{eqn:g_fun}
\end{eqnarray}
where $f_{\theta}(\bullet|\mathbf{s})$ is defined in equation \eqref{eqn:Condmassfunc}.
We define the \emph{$k$-component parametric mixture model ($k$-component
PMM) }as 
\begin{eqnarray}
g(\bullet|\mathbf{s};\theta_{0},\ldots,\theta_{k-1},\pi_{0},\ldots,\pi_{k-1}) & = & \sum_{i=0}^{k-1}\pi_{j}g_{\theta_{j}}(\bullet|\mathbf{s})\label{eqn:PMM}
\end{eqnarray}
 where $\theta_{0}=0$ and $\theta_{j}\ne\theta_{J}$, if $j\ne J$. 

Let $\mathbf{T}=\left\langle T_{1},T_{2},\cdots,T_{m}\right\rangle $
and $\mathbf{t}=\left\langle t_{1},t_{2},\cdots,t_{m}\right\rangle $
be vectors of the $T_{i}$s and $t_{i}$s used in equation \eqref{eqn:BF}.
Assuming $T_{i}$ is independent of $T_{j}$ and $S_{j}$ for any
$i\ne j$, the joint probability mass function is 
\begin{eqnarray}
g(\mathbf{t}|\mathbf{s};\theta_{0},\ldots,\theta_{k-1},\pi_{0},\ldots,\pi_{k-1}) & = & \prod_{i=1}^{m}g(t_{i}|\mathbf{s};\theta_{0},\ldots,\theta_{k-1},\pi_{0},\ldots,\pi_{k-1})\label{eqn:joint_prob}\\
 & = & \prod_{i=1}^{m}g(t_{i}|s_{i};\theta_{0},\ldots,\theta_{k-1},\pi_{0},\ldots,\pi_{k-1})\nonumber 
\end{eqnarray}
where $s_{i}$ is the observed value of $S_{i}$ for GO category $i$,
and $\mathbf{s}=\left\langle s_{1},s_{2},\cdots,s_{m}\right\rangle $.

Moreover, we assume that for given the number of genes in GO category
$i$, $T_{i}$ $(i=1,\ldots,m)$, satisfies the $k$-component PMM
shown in equation \eqref{eqn:PMM}. In other words, we assume that
the possible log odds ratios of GO category $i$ are the $\theta_{0},\theta_{1},\theta_{2},\ldots,\theta_{k-1}$
of equation \eqref{eqn:PMM} if the alternative hypothesis $H_{1}$
in the hypothesis comparison \eqref{eqn:two-sided-hypothesis-GO}
is true. 

Therefore, the log-likelihood function under the $k$-component PMM
for all GO categories is 
\begin{eqnarray}
\log L(\theta_{0},\ldots,\theta_{k-1},\pi_{0},\ldots,\pi_{k-1}) & = & \log g(\mathbf{t}|\mathbf{s};\theta_{0},\ldots,\theta_{k-1},\pi_{0},\ldots,\pi_{k-1})\nonumber \\
 & \text{=} & \sum_{i=1}^{m}\left[\log\sum_{j=0}^{k-1}\pi_{j}g_{\theta_{j}}(t_{i}|s_{i})\right]\label{eqn:loglikPMM}
\end{eqnarray}
The LFDR of GO category $i$ is estimated by 
\begin{equation}
\widehat{\text{LFDR}}_{i}^{\left(k\right)}=\frac{\widehat{\pi}_{0}g_{\theta_{0}}(t_{i}|s_{i})}{g(t_{i}|s_{i};\theta_{0},\widehat{\theta}_{1},\ldots,\widehat{\theta}_{k-1},\widehat{\pi}_{0},\ldots,\widehat{\pi}_{k-1})}\label{eqn:LFDR_typeII}
\end{equation}
where $\widehat{\theta}_{1},\ldots,\widehat{\theta}_{k-1}$ and $\widehat{\pi}_{0},\ldots,\widehat{\pi}_{k-1}$
are maximum likelihood estimates of $\theta_{1},\ldots,\theta_{k-1}$
and $\pi_{0},\ldots,\pi_{k-1}$ in equation \eqref{eqn:loglikPMM}.
We call $\widehat{\text{LFDR}}_{i}^{\left(k\right)}$ the $k$\emph{-component
maximum likelihood estimator} (MLE$k$).

\subsection*{LFDR estimator based on the normalized maximum likelihood\label{sec:NMLE}}

Combining equations \eqref{eqn:OR}-\eqref{eqn:LFDR-computation},
we obtain

\begin{equation}
\text{LFDR}_{i}=\left(1+\text{BF}_{i}\times\frac{(1-\pi_{0})}{\pi_{0}}\right)^{-1}\label{eqn:Bayes}
\end{equation}
Therefore, given a guessed value of $\pi_{0}$, we may use an estimator
of the Bayes factor to estimate the LFDR of a GO category. 

We next develop such an estimator of the Bayes factor. For GO category
$i$, let $\mathcal{E}_{i}$ stand for the set of all probability
mass functions defined on $\left\{ 0,1,\dots,s_{i}\right\} $, the
set of all possible values of $t_{i}$. Based on the hypothesis comparison
\eqref{eqn:two-sided-hypothesis-GO}, the set of log odds ratios,
denoted as $\Theta$, is $\{0\}$ under the null hypothesis and is
the set of all real values except $0$ under the alternative hypothesis.
With the assumption that the random variable $T_{i}$ is independent
of the random variable $S_{j}$ for any $i\ne j$, the \emph{regret}
of a predictive mass function $\bar{f}\in\mathcal{E}_{i}$ is a measure
of how well it predicts the observed value $t_{i}\in\left\{ 0,1,\ldots,s_{i}\right\} $.
The regret is defined as 
\begin{equation}
\text{reg}(\bar{f},t_{i}|s_{i};\Theta)=\log\frac{f_{\hat{\theta}_{i}(t_{i}|s_{i})}(t_{i}|s_{i})}{\bar{f}(t_{i}|s_{i})}\label{eqn:regret}
\end{equation}
where $\hat{\theta}_{i}(t_{i}|s_{i})$ is the Type I MLE with respect
to the $\Theta$ under the observed values $t_{i}$ given $s_{i}$
\citep{NMWL,RefWorks:375}. 

For all members of $\mathcal{E}_{i}$, the \emph{optimal predictive
conditional probability mass function} of GO category $i$, denoted
as $f_{i}^{\dagger}$, minimizes the maximal regret in the sample
space $\left\{ 0,1,\ldots,s_{i}\right\} $ in the sense that it satisfies
\begin{equation}
f_{i}^{\dagger}=\arg\min_{\bar{f}\in\mathcal{E}_{i}}\max_{t\in\left\{ 0,1,\ldots,s_{i}\right\} }\text{reg}(\bar{f},t|s_{i};\Theta)\label{eqn:minimax}
\end{equation}
It is well known \citep{RefWorks:375} that the predictive probability
mass function that satisfies equation \eqref{eqn:minimax} is 

\begin{eqnarray}
f_{i}^{\dagger}(t_{i}|s_{i};\Theta) & = & \frac{\max_{\theta\in\Theta}f_{\theta}(t_{i}|s_{i})}{\mathcal{K}_{i}^{\dagger}(\Theta)}\label{eqa:NMCL}
\end{eqnarray}
where $f_{\theta}(t_{i}|s_{i})$ is the conditional probability mass
function defined in equation \eqref{eqn:Condmassfunc}, and $\mathcal{K}_{i}^{\dagger}(\Theta)$
is the constant defined as 
\begin{eqnarray}
\mathcal{K}_{i}^{\dagger}(\Theta) & = & \max_{\theta\in\Theta}f_{\theta}(y|s_{i})\label{eqa:COMPNMCL}\\
 & = & \sum_{y=\max(0,s_{i}-n_{2})}^{\min(s_{i},n_{1})}\max_{\theta\in\Theta}f_{\theta}(y|s_{i})\nonumber \\
 & = & \sum_{y=\max(0,s_{i}-n_{2})}^{\min(s_{i},n_{1})}\frac{\binom{n_{1}}{y}\binom{n_{2}}{s_{i}-y}e^{y\hat{\theta}_{i}(y)}}{\sum_{j=\max(0,s_{i}-n_{2})}^{\min(s,n_{1})}\binom{n_{1}}{j}\binom{n_{2}}{s_{i}-j}e^{j\hat{\theta}_{i}(y)}}\nonumber 
\end{eqnarray}
where 
\begin{equation}
\hat{\theta}_{i}(y)=\arg\max_{\theta\in\Theta}f_{\theta}(y|s_{i})\label{eqn.condmle}
\end{equation}
We call $f_{i}^{\dagger}(t_{i}|s_{i};\Theta)$ the \emph{normalized
maximum likelihood} (NML) associated with the hypothesis that $\theta_{i}\in\Theta$.

Thus, $\text{BF}_{i}$ is estimated by 
\begin{equation}
\widehat{\text{BF}}_{i}^{\dagger}=\frac{f_{i}^{\dagger}(t_{i}|s_{i};{\theta:\theta\ne0})}{f_{i}^{\dagger}(t_{i}|s_{i};{0})},\label{eqn:BF_hat_NML}
\end{equation}
which we call the the \emph{NML ratio}. Therefore, by combining equations
\eqref{eqn:BF} and \eqref{eqn:OR}, if we guess the prior probability
$\pi_{0}$, the LFDR estimate of GO category $i$ in the hypothesis
comparison \eqref{eqn:two-sided-hypothesis-GO} is 

\begin{equation}
\widehat{\text{LFDR}}_{i}^{\dagger}=\left[1+\frac{1-\pi_{0}}{\pi_{0}}\times\widehat{\text{BF}}_{i}^{\dagger}\right]^{-1}\label{eqn:LFDR_Bayes_Twosided}
\end{equation}
where $\widehat{\text{BF}}_{i}^{\dagger}$ is defined in equation
\eqref{eqn:BF_hat_NML}. We call this LFDR estimator the \emph{normalized
maximum likelihood estimator} (NMLE).

To assess the performance of the NML ratio $\widehat{\text{BF}}_{i}^{\dagger}$,
it will be compared to the following estimate of the Bayes factor.
Equations \eqref{eqn:Bayes} and \eqref{eqn:LFDR_typeII} suggest
\begin{equation}
\widehat{\text{BF}}_{i}=\frac{1-\widehat{\text{LFDR}}_{i}^{\left(k\right)}}{\widehat{\text{LFDR}}_{i}^{\left(k\right)}}\times\frac{1-\widehat{\pi}_{0}}{\widehat{\pi}_{0}}\label{eqn:BF_hat_MLE}
\end{equation}
as an MLE-based estimator of $\text{BF}_{i}$.

\section*{Results \label{sec:Results}}

\subsection*{Breast cancer data analysis\label{sub:Real-data-analysis}}

The data set used here is from an experiment applying an estrogen
treatment to cells of a human breast cancer cell line \citep{scholtens_analyzing_2004}.
The data, which is available from the Bioconductor project, contains
$8$ Affymetrix HG-U95Av2 CEL files from an estrogen receptor-positive
breast cancer cell line. (For further information concerning the data
and also the Bioconductor project, see \citet{Gentleman2005}.) For
simplicity of terminology, we consider probes in the microarray experiment
as genes, and use the $12,625$ genes expressed in the microarray
experiment as a reference.

We selected as genes of interest those that were differentially expressed
between two groups according to the following criterion. Using the
LFDR as the probability that a gene is EE, we considered genes with
LFDR estimates below $0.2$ as DE. In other words, we selected as
DE genes those were differentially expressed with estimated posterior
probability of at least $80\%$. We used the 2-sample t-test with
equal variances to compute the p-value of each gene in the microarray.
The LFDR of every gene is estimated using the theoretical null hypothesis
method of \citet{RefWorks:55,efron_large-scale_2010}; empirical null
hypotheses can lead to excessive bias due to deviations from normality
\citep{conditional2009}. When we compared gene expression data for
the presence and absence of estrogen after $10$ hours of exposure,
we obtained $74$ DE genes. 

\begin{figure}[H]
\begin{centering}
\includegraphics[scale=0.85]{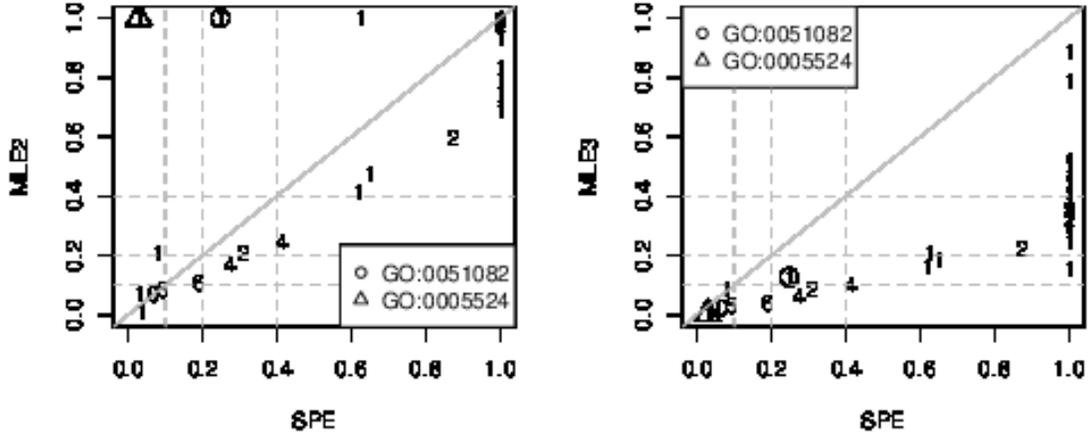}
\par\end{centering}

\caption{Comparison of the LFDR estimated by the SPE with the LFDR estimated
by the MLE2 (left) and MLE3 (right). Each integer represents a number
of GO categories. $\text{Intergers}>1$ indicate ties. \label{fig:PLFDR_vs_TLFDR}}
\end{figure}

\begin{figure}[h]
\begin{centering}
\includegraphics[scale=0.85]{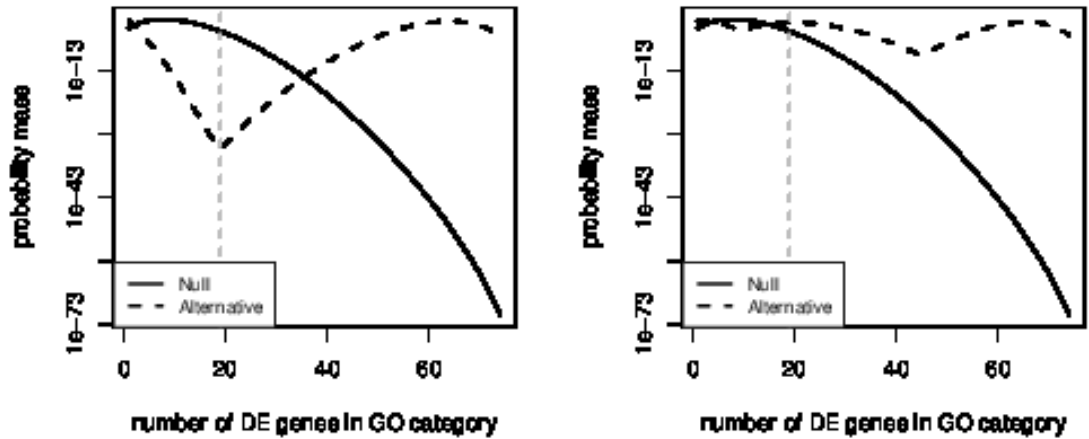}
\par\end{centering}

\caption{The conditional probability mass functions given the number of genes
in $\text{GO:}0005524$ under a null hypothesis, and alternative hypotheses
based on the $2$-component PMM (left) and $3$-component PMM (right).
The grey dashed line is the number of DE genes in $\text{GO:}0005524$.
\label{fig:log_PMF_GO0005524}}
\end{figure}

\begin{figure}
\begin{centering}
\includegraphics[scale=0.85]{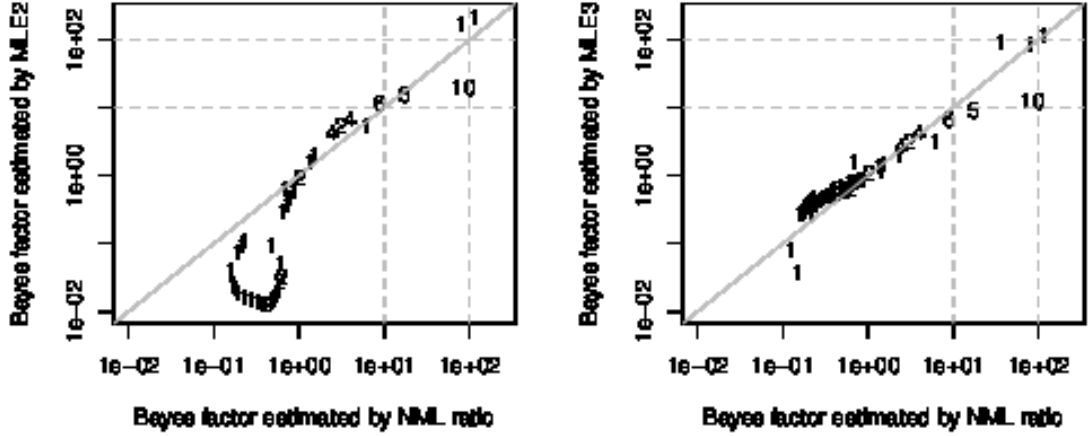}
\par\end{centering}

\caption{Comparison of the Bayes factor estimated by the NML ratio with that
estimated by the MLE2 (left) and MLE3 (right). The integers are defined
in Figure \ref{fig:PLFDR_vs_TLFDR}. The grey dashed lines mark commonly
used thresholds for strong and overwhelming evidence \citep{RefWorks:182,RefWorks:435}.\label{fig:BF_NMCL_TypeII}}
\end{figure}

Defining \emph{unrelated} pairs of GO categories as those that do
not share any common ancestor, we selected for analysis all unrelated
GO molecular function categories with at least $1$ DE gene, thereby
obtaining a total of $82$ GO categories of interest. For each GO
category, the p-value used in SPE to estimate LFDR is computed based
on the 2-sided Fisher's exact test. Figure \ref{fig:PLFDR_vs_TLFDR}
compares the SPE to the MLEs based on the $2$-component (MLE2) and
$3$-component (MLE3) PMM. Figure \ref{fig:log_PMF_GO0005524} displays
the probability mass of $\text{GO:}0005524$ under the null and alternative
hypotheses of the hypothesis comparison \eqref{eqn:two-sided-hypothesis-GO}.
Figure \ref{fig:BF_NMCL_TypeII} compares MLE-based estimates of the
Bayes factor given by equation \eqref{eqn:BF_hat_MLE} to the NML
ratios given by equation \eqref{eqn:BF_hat_MLE}.

\subsection*{Simulation studies\label{sub:Simulations}}

The aim of the following simulation studies is to compare the LFDR
estimation bias of SPE, MLE2, and MLE3. The NMLE is not taken into
account because its performance depends not only on the data, but
also on the specified prior probability $\pi_{0}$.

The simulation setting involves $10,000$ genes in a microarray with
$200$ genes identified as DE and $100$ GO categories. We conducted
a separate simulation study using each of these values of $\pi_{0}$:
$50\%$, $60\%$, $70\%$, $80\%$, $90\%$, and $94\%$.

Since the PMM behind the MLE is optimal when the number of GO categories
with overrepresentation ({}``enrichment'') is equal to the number
with underrepresentation ({}``depletion''), we assessed the sensitivity
of the MLE to that symmetry assumption by using strongly asymmetric
log odds ratios as well as those that are symmetric. For each GO category,
two configurations were used in this simulation to choose log odds
ratios: the \emph{asymmetric configuration} shown in equation \eqref{eqn:pos-LOR}
and the \emph{symmetric configuration} shown in equation \eqref{eqn:sym-LOR}. 

\begin{equation}
\theta_{i}^{\text{asymmetric}}=\begin{cases}
\frac{5i}{100(1-\pi_{0})}, & 1\le i\le100(1-\pi_{0})\\
0, & 100(1-\pi_{0})<i\le100
\end{cases}\label{eqn:pos-LOR}
\end{equation}

\begin{equation}
\theta_{i}^{\text{symmetric}}=\begin{cases}
\frac{i}{10(1-\pi_{0})}, & 1\le i\le50(1-\pi_{0})\\
5-\frac{i}{10(1-\pi_{0})}, & 50(1-\pi_{0})<i\le100(1-\pi_{0})\\
0, & 100(1-\pi_{0})<i\le100
\end{cases}\label{eqn:sym-LOR}
\end{equation}

Considering the log odds ratios of all $100$ GO categories constructed
by either the asymmetric or the symmetric configuration, we generated
Table \ref{tab:twobytwogenes} for each GO category as follows: 
\begin{itemize}
\item $x_{1}$ is generated from a binomial distribution with the parameter
$\Pi_{1}$ used in equation \eqref{eqa:FullLikelihood}; $\Pi_{1}$
is a real value randomly picked from $0$ to $1$ .
\item $x_{2}$ is obtained from a binomial distribution with the parameter
$\Pi_{2}=\left[\frac{(1-\Pi_{1})\times2^{\theta_{i}}}{\Pi_{1}}+1\right]^{-1}$,
obtained by solving equation \eqref{eqn:theta}. 
\end{itemize}
The p-value of each GO category used in the SPE is obtained from 2-sided
Fisher's exact test. The $k$-component PMM ($k=2\text{ or }k=3$)
used in the MLE is shown in equation \eqref{eqn:PMM} with $\pi_{j}=\left(1-\pi_{0}\right)/k\,\left[j=1,\dots,k\right]$
and $g_{\theta_{i}}(\bullet|s)=g_{\theta_{i}}(t_{i}|s_{i})$ defined
in equation \eqref{eqn:g_fun}. For every log odd ratio sequence,
we estimated the LFDR $20$ times using the SPE, MLE2, and MLE3. We
compared the performances of the $3$ estimators by means of estimating
the LFDR bias. The true LFDR is computed by equation \eqref{eqn:LFDR-computation},
where 
\[
f_{0}(t_{i})=\frac{\binom{n}{t_{i}}\binom{N-n}{s_{i}-t_{i}}}{\sum_{j=\max(0,s_{i}+n-N)}^{\min(s_{i},n)}\binom{n}{j}\binom{N-n}{s_{i}-j}}
\]
 and $f_{1}(t_{i})$ is computed by 
\[
\frac{1}{J}\sum_{j=1}^{J}f_{\theta_{j}}(t_{i}|s_{i})
\]
where $f_{\theta}(t|s)$ is defined in equation \eqref{eqn:Condmassfunc}.

\begin{figure}
\begin{centering}
\includegraphics[scale=0.76]{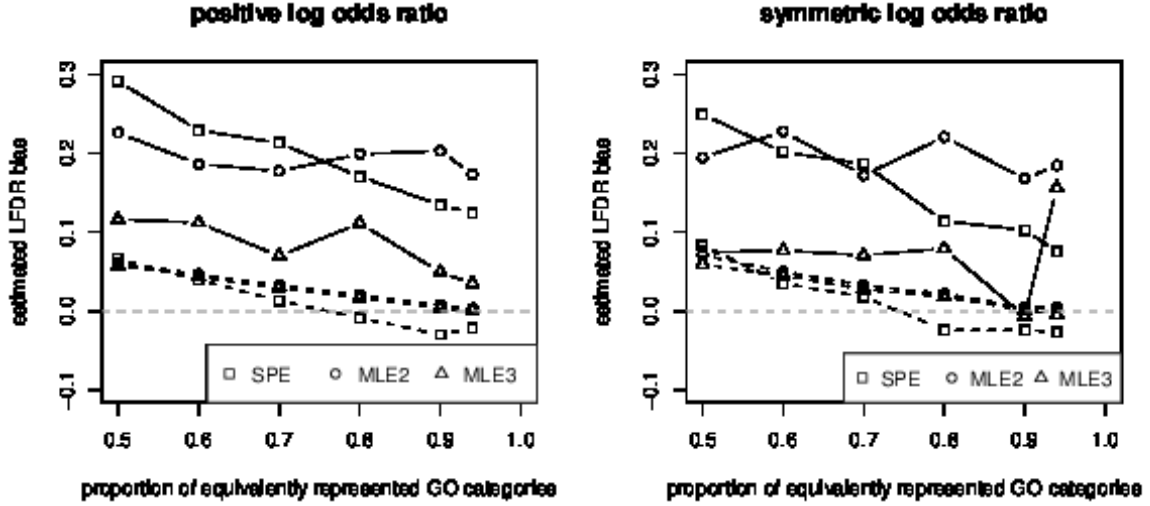}
\par\end{centering}

\caption{The performance of LFDR estimators for equivalently (dashed line)
or differentially (solid line) represented GO categories.\label{fig:estimators-performance}}
\end{figure}

Figure \ref{fig:estimators-performance} shows the performance comparisons
of the $3$ LFDR estimators for simulation data obtained from the
symmetric and asymmetric log odds ratios. The LFDR biases estimated
by the SPE and MLE2 are similar. The LFDR estimated by the MLE3 provides
the lowest bias among the $3$ LFDR estimators. Moreover, the estimated
LFDR biases of the estimators are not strongly affected by whether
the log odds ratios are symmetric or asymmetric. Furthermore, the
bias of the LFDR estimated by the SPE decreases as $\pi_{0}$, the
probability that GO categories are equivalently represented, increases.
However, the LFDR estimate attains a negative bias if $\pi_{0}$ is
higher than $80\%$. In other words, some equivalently represented
GO categories are declared as differentially represented GO categories. 

\section*{Conclusions\label{sec:Conclusions}}

Efron's method \citep{RefWorks:55,efron_large-scale_2010} can be
used to estimate GO categories and thus address the gene enrichment
problem, provided that thousands of GO categories are taken into account.
However, in most gene enrichment studies, researchers focus on medium-
or small-scale numbers of GO categories, i.e., several hundred, dozens
or only one GO category. Here, we adapted $3$ LFDR estimators (the
SPE, MLE, and NMLE) to address the gene enrichment problem with medium-
and small-scale numbers of GO categories, and compared these using
breast cancer and simulation data. 

The MLE is sensitive to $k$, the number of PMM components. The MLE
is used when considering a medium-scale number of GO categories, i.e.,
$100$. In our breast cancer data analysis, the estimated LFDRs of
$\text{GO:}0051082$ and $\text{GO:}0005524$ using MLE2 were $100\%$
(Figure \ref{fig:PLFDR_vs_TLFDR}). However, the LFDRs estimated by
MLE3 were very close to 0. Using the MLE formula shown in equation
\eqref{eqn:LFDR_typeII}, and the $k$-component PMM shown in equation
\eqref{eqn:PMM}, we determined that the sensitivity of the LFDRs
of GO category $i$ estimated by MLE2 and MLE3 depended mainly on
the sensitivity of the Bayes factor, based on the number of PMM components.
Comparing the probability masses of $\text{GO:}0005524$, based on
the $2$- and $3$-component PMMs shown in Figure \ref{fig:log_PMF_GO0005524},
we found that the probability mass of $\text{GO:}0005524$ under the
null hypothesis is larger than that under the alternative hypothesis
based on the $2$-component PMM (left plot in Figure \ref{fig:log_PMF_GO0005524}).
By contrast, the probability mass under the null hypothesis is smaller
than that under the alternative hypothesis based on the $3$-component
PMM (right plot in Figure \ref{fig:log_PMF_GO0005524}). Thus, the
LFDR estimated by the MLE is strongly dependent on the number of PMM
components.

Nevertheless, the performance comparison in Figure \ref{fig:estimators-performance}
indicates that the MLE has lower bias than the SPE when the number
of GO categories is much larger than $k$ even when the ideal value
of $k$ is unknown. Moreover, MLE3 has lower bias than MLE2 as an
LFDR estimator. However, when the number of GO categories is not much
larger than $k$, the estimated proportion of GO categories equivalently
represented become strongly biased toward $0$. In that situation,
the false positive rate increases as the number of PMM components. 

Due to its conservatism and freedom from the PMM, we recommend using
the SPE when the number of GO categories of interest is too small
for the MLE, e.g., about $10$ categories. Based on the simulations
reported by \citet{BFDR}, we conjecture that the SPE has acceptably
low LFDR-estimation bias when there are at least 3 GO categories.

Finally, we recommend that the NMLE be used given only 1 or 2 GO categories
of interest. Neither the MLE nor the SPE is able to estimate the LFDR
for only $1$ GO category of interest; moreover, they probably have
excessive bias when based on only $2$ GO categories. Thus, the NMLE
is the recommended method of addressing the gene enrichment problem
in this smallest-scale case. The NMLE depends not only on the data
but also on a guess of the value of $\pi_{0}$, which, in the absence
of strong prior information, is often set to the default value of
50\%. A closely related approach is to use the NML ratio as an estimate
of the Bayes factor directly without guessing $\pi_{0}$. By using
10 and 100 as thresholds of each estimated Bayes factor to determine
whether a GO category is differentially represented, we reached similar
conclusions whether using the NML and or an MLE (Figure \ref{fig:BF_NMCL_TypeII}).
Thus, at least for our data set, the NML ratio tends to estimate the
Bayes factor almost as accurately as methods that simultaneously use
information across GO terms.

\section*{Acknowledgments\label{sec:Acknowledgments}}

We thank both Editage and Donna Reeder for detailed copyediting. We
are grateful to Corey Yanofsky and Ye Yang for useful discussions.
This work was partially supported by the Natural Sciences and Engineering
Research Council of Canada, by the Canada Foundation for Innovation,
by the Ministry of Research and Innovation of Ontario, and by the
Faculty of Medicine of the University of Ottawa.

\section*{}

\bibliographystyle{plainnat}
\bibliography{JerryRefs,refman}

\end{document}